\documentstyle[epsfig]{rspublic}

\newcommand{\Lya}{Ly$\alpha\ $}

\newcommand{\emunits}{{\rm\,ergs\,Gpc^{-3}\,s^{-1}\,Hz^{-1}}}

\newcommand{\etal}{et~al.\ }
\def\kms{\,{\rm km\,s^{-1}}}
\def\kmsmpc{\,{\rm km\,s^{-1}\,Mpc^{-1}}}
\def\msun{\,{\rm M_\odot}}

\def\sfrd{\,{\rm M_\odot\,yr^{-1}\,Mpc^{-3}}}
\def\spose#1{\hbox to 0pt{#1\hss}}
\def\lta{\mathrel{\spose{\lower 3pt\hbox{$\mathchar"218$}} \raise 2.0pt\hbox{$\mathchar"13C$}}}
\def\gta{\mathrel{\spose{\lower 3pt\hbox{$\mathchar"218$}} \raise 2.0pt\hbox{$\mathchar"13E$}}}

\def\lya{Ly$\alpha\ $}

\def\emunits{\,{\rm ergs\,s^{-1}\,Hz^{-1}\,Mpc^{-3}}}
\def\ndotunits{\,{\rm s^{-1}\,Mpc^{-3}}}
\def\ndotun{\,{\rm phot\,s^{-1}}}
 \def\ni{\noindent}
\def\HI{\hbox{H~$\scriptstyle\rm I\ $}}
\def\HII{\hbox{H~$\scriptstyle\rm II\ $}}
\def\HeI{\hbox{He~$\scriptstyle\rm I\ $}}
\def\HeII{\hbox{He~$\scriptstyle\rm II\ $}}
\def\HeIII{\hbox{He~$\scriptstyle\rm III\ $}}

\def\nH{{\rm H}}

\def\nHe{{\rm He}}

\def\nHeIII{{\rm HeIII}}
\def\ni{{\noindent}}
\def\AB{{\rm AB}}

\begin{document}

\title[Reionization]{\bf Cosmological Reionization
}

\author[P. Madau]{Piero Madau}

\affiliation{Institute of Astronomy, University of Cambridge\\
Madingley Road, CB3 0HA, Cambridge, UK}

\label{firstpage}

\maketitle

\begin{abstract}{diffuse radiation -- intergalactic medium -- radiative 
transfer}
\ni In popular cosmological scenarios, some time beyond a redshift of 10,
stars within protogalaxies created the first heavy elements; these 
systems, together 
perhaps with an early population of quasars, generated the ultraviolet 
radiation and mechanical energy that reheated and reionized the cosmos.
The history of the Universe during and soon after these crucial formative
stages is recorded in the all-pervading intergalactic medium (IGM), which
contains most of the ordinary baryonic
material left over from the big bang. Throughout the epoch of structure
formation, the IGM becomes clumpy and acquires peculiar motions under the
influence of gravity, and acts as a source for the gas that gets
accreted, cools, and forms stars within galaxies, and as a sink for the
metal enriched material, energy, and radiation which they eject.

\end{abstract}

\section{Introduction}
\ni At epochs corresponding to $z\sim 1000$ the intergalactic medium (IGM) 
is expected to recombine and remain neutral until sources of radiation and 
heat develop that are capable of
reionizing it. The detection of transmitted flux shortward of the \lya
wavelength in the spectra of sources at $z\sim 5$ implies that the hydrogen
component of this IGM was ionized at even higher redshifts.
There is some evidence that the  double reionization of helium may have
occurred later, but this is still controversial.
It appears then that substantial sources of ultraviolet photons and mechanical
energy were already
present when the Universe was less than 7\% of its current age, perhaps
quasars and/or young star-forming galaxies:
an episode of pre-galactic star formation may provide a possible explanation
for the widespread existence of heavy elements (like carbon, oxygen, and
silicon) in the IGM, while the integrated radiation emitted from quasars is 
likely responsible for the reionization of the intergalactic helium.
Establishing the epoch of reionization and reheating is crucial for determining
its impact on several key cosmological issues, from
the role reionization plays in allowing protogalactic objects to cool and
make stars, to determining the small-scale structure in the temperature
fluctuations of the cosmic microwave background. Conversely, probing the
reionization epoch may provide a means for constraining competing models for
the formation of cosmic structures, and of detecting the onset of the first
generation of stars, galaxies, and black holes in the Universe.

\section{The transition from a neutral to an ionized Universe}
\ni Popular cosmological models predict that most of the intergalactic hydrogen
was reionized by the first generation of stars or accreting black holes at 
$z=7-15$. One should note, however, that while numerical 
N-body$+$hydrodynamical simulations
have convincingly shown that the IGM is expected to fragment into structures at
early times in cold dark matter (CDM) cosmogonies (e.g. Cen \etal 1994;
Zhang, Anninos, \& Norman 1995; Hernquist \etal 1996),
the same simulations are much less able to predict the efficiency with
which the first gravitationally collapsed objects lit up the Universe
at the end of the `dark age' (Rees, this volume). 

\subsection{Photo- versus collisional ionization}

\ni The scenario that has received the most theoretical studies is one 
where hydrogen is photoionized by the UV radiation emitted either by quasars or
by  stars with masses $\gta 10\,\msun$, rather than ionized by collisions with 
electrons heated up by, e.g. supernova-driven winds from early pregalactic
(`Pop III') objects.  
In the former case a high degree of ionization requires
about $13.6\times (1+t/\bar t_{\rm rec})\,$eV per hydrogen atom, where 
$\bar t_{\rm rec}$ is the volume-averaged hydrogen recombination timescale, 
$t/\bar t_{\rm rec}$ being much greater than unity already at 
$z\approx 10$ according to the numerical simulations of Gnedin \& Ostriker
(1997), and Gnedin (2000). Collisional ionization to a neutral fraction of only
few parts in $10^{5}$ requires a comparable energy input, i. e. an IGM 
temperature close to $10^5\,$K or about $25\,$eV per atom. 

Massive stars will deposit both radiative and mechanical energy into the 
interstellar medium of Pop III objects. A complex network of `feedback' 
mechanisms is likely at work in these systems, as the gas in shallow potential
is more easily blown away thereby quenching further star formation (Mac Low \& 
Ferrara 1999), and the blastwaves produced by supernova explosions reheat 
the surrounding intergalactic gas and enrich it with newly formed heavy 
elements and dust. It is therefore difficult to establish whether an early 
input of mechanical energy will actually play a major role in determining the 
thermal and ionization state of the IGM on large scales (Tegmark, Silk, \& 
Evrard 1993). What can be easily shown is that, during the evolution of a a 
`typical' 
stellar population, more energy is lost in ultraviolet radiation than in
mechanical form. This is because in nuclear burning from zero to solar 
metallicity ($Z_\odot=0.02$), the energy radiated per 
baryon is $0.02\times 0.007\times m_\nH c^2$; about one third of it goes into 
H-ionizing photons. The same massive stars that dominate the UV light 
also explode as supernovae (SNe), returning most of the metals to the 
interstellar medium and 
injecting about $10^{51}\,$ergs per event in kinetic energy. For a Salpeter 
initial mass function (IMF), one has about one SN every $150\,\msun$ of baryons
that forms stars. The mass fraction in mechanical energy is then approximately
$4\times 10^{-6}$, ten times lower than the fraction released in photons
above 1 ryd. 

The relative importance of photoionization versus shock ionization will 
depend, however, on the efficiency with which radiation and mechanical 
energy actually escape into the IGM.    
Consider, for example, the case of an early generation of halos with circular 
speed $v_c=50\,\kms$, corresponding in top-hat spherical collapse to a virial 
temperature $T_v=0.5\mu m_p v_c^2/k\approx 10^{5.3}\,$K
and halo mass $M=0.1v_c^3/GH\approx 10^9 [(1+z)/10]^{-3/2}
h^{-1}\, \msun$.\footnote{This assumes an Einstein-de Sitter (EdS) Universe 
with $H_0=100\,h\,\kmsmpc$.}~ In these systems rapid cooling by
atomic hydrogen can take place and a significant fraction,
$f\Omega_B$, of their total mass may be converted into stars over a 
dynamical timescale (here $\Omega_B$ is the baryon density parameter). For 
$f=0.05$, $\Omega_Bh^2=0.02$, and $h=0.5$, the 
explosive output of $50,000$ SNe would inject an energy $E_0\approx 
10^{55.7}\,$ergs. The hot gas will
escape its host, shock the IGM, and eventually form a cosmological blast wave.
If the explosion occurs at cosmic time $t=4\times 10^8\,$yr, corresponding in 
the adopted cosmology (EdS with $h=0.5$) to $z=9$, at time $\Delta t=0.2t$ 
after the event the proper radius of the (adiabatic) shock is given by the 
standard 
Sedov-Taylor self-similar solution,
$$ 
R_s\approx \left(\frac{12\pi G E_0}{\Omega_B}\right)^{1/5}t^{2/5}\Delta t^{2/5}
\approx 23\,{\rm kpc}.  \eqno(1)  
$$
At this instant the shock velocity relative to the Hubble flow is  
$$
v_s\approx 2R_s/5\Delta t\approx 110\,\kms,  \eqno(2)  
$$
still much higher than the escape velocity from the halo. The gas temperature 
just behind the shock front is $T_s=3\mu m_pv_s^2/16k 
\approx 4\times 10^5\,$K, more than enough to efficiently ionize all the 
incoming hydrogen. At these redshifts, it is the onset of Compton cooling off 
cosmic microwave background photons that ends the adiabatic stage of blast 
wave propagation. According to the Press-Schechter formalism, the 
comoving abundance of collapsed dark halos with mass $M=10^9\,h^{-1}\,M_\odot$ 
at $z=9$ is $dn/d\ln M\sim 5\,h^3\,$Mpc$^{-3}$, corresponding to a mean 
proper distance between neighboring halos of $\sim 40\,h^{-1}\,$kpc, and to a 
total mass density parameter of order $0.02$. With the 
assumed star formation efficiency, only a small fraction, about one 
percent, of the stars seen today would have to be formed at these early 
epochs. Still, our simple analysis shows that the blast waves from such 
a population of pregalactic objects could overlap with large enough velocities 
to initially drive the intergalalactic medium to a significantly higher 
adiabat, $T\gta 10^5\,$K, than expected from photoionization, and pollute the 
entire IGM with metal-enriched material. A lower density of sources -- which 
would therefore have to originate from higher amplitude peaks -- would suffice 
if the typical efficiency of star formation were larger than assumed here. 

Quasar-driven blast waves ($E_0\gta 10^{60}\,$ergs) are instead quite 
inefficient at ionizing the IGM, since much of the initial 
explosion energy is lost into the collisionless component (Voit 1996). 
They would also be too rare to fill 
the IGM without violating the COBE limit on the $y$-distortion of the microwave 
background. 

\subsection{Cosmological \HII regions}

\ni In the following sections we will focus our attention to the  
photoionization of the IGM, i.e. we will assume that UV
photons from an early generation of stars and/or quasars are the main source
of energy for the reionization and reheating of the Universe, and that 
star formation and quasar activity occurs in collapsed galaxy halos.  
The process then begins as individual sources start to generate
expanding \HII regions in the surrounding IGM; throughout an \HII region, H is
ionized and He is either singly or doubly ionized. As more and more sources of
ultraviolet radiation switch on, the ionized volume grows in size while the 
neutral phase shrinks. Reionization is completed when the \HII regions 
overlap, and every point in the intergalactic space gets exposed for the first 
time to a nearly uniform Lyman-continuum (Lyc) background.

When an isolated point source of ionizing radiation turns on, an ionization 
(I) front separating the \HII and \HI regions propagates into the neutral gas, 
and the proper volume $V_I$ of the ionized zone grows according to the equation
$$
\frac{dV_I}{dt}-3HV_I=\frac{\dot N_{\rm ion}}{\bar{n}_\nH}-\frac{V_I}
{\bar{t}_{\rm rec}}, \eqno(3)
$$
(Shapiro \& Giroux 1987), where $\dot N_{\rm ion}$ is the number of ionizing 
photons emitted by the 
central source per unit time, $\bar{n}_\nH(0)=1.7\times 10^{-7}$ $(\Omega_B 
h^2/0.02)$ cm$^{-3}$ is today's mean hydrogen density, and all other symbols 
have their usual meaning. Most photons travel freely in the ionized gas, 
and are absorbed in a transition layer. 
In the case of stellar sources the I-front is quite sharp, and the degree of
ionization changes on a short distance of the order of the mean free path for 
an ionizing photon. When $\bar{t}_{\rm rec}\ll t$, the growth of the
\HII region is
slowed down by recombinations in the highly inhomogeneous IGM, and its evolution
can be decoupled from the Hubble expansion. Just like in the static 
case, the ionized bubble then fills its time-varying Str\"omgren sphere
after a few recombination timescales,
$$
V_I=\frac{\dot N_{\rm ion}\bar{t}_{\rm rec}}{\bar{n}_\nH}
(1-e^{-t/\bar{t}_{\rm rec}}). \eqno(4)
$$
While the volume that is ionized depends on the luminosity of the
central source, the time it takes to produce an ionization-bounded
region is only a function of $\bar{t}_{\rm rec}$. 

In the presence of a population of ionizing sources, the transition from a
neutral IGM to one that is almost fully ionized can be statistically described 
by the evolution with redshift of the {\it volume filling factor} (or porosity)
$Q$ of \HII, \HeII, and \HeIII regions. The radiation emitted by spatially
clustered stellar-like and quasar-like sources -- the number densities and
luminosities of which may change rapidly as a function of redshift --
coupled with absorption processes in a medium that becomes more and more 
clumpy owing to the non-linear collapse of structures (Figure 1), all
determine the complex topology of neutral and ionized zones in the Universe
(Gnedin 2000; Ciardi \etal 2000; Abel, Norman, \& Madau 1999).
When $Q\ll1$ and the radiation sources are randomly distributed, the ionized
regions are spatially isolated, every UV photon is absorbed somewhere in the
IGM, and the UV radiation field is highly inhomogeneous. As $Q$ grows, the 
crossing of ionization fronts becomes more and more common, until 
percolation occurs at $Q=1$.

\begin{figure}
\epsfysize=7cm 
\epsfxsize=7cm 
\hspace{3.5cm}\epsfbox{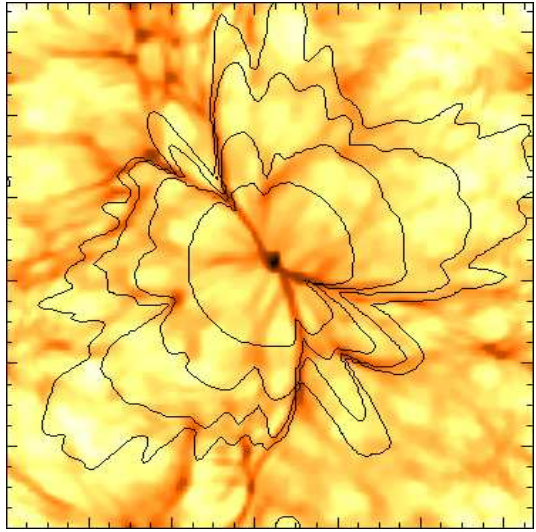}
\vspace{0.5cm}
\caption{Simulating the reionization of the Universe:
propagation of an ionization front in a $128^3$
cosmological density field. A `mini-quasar' emitting $5\times
10^{53}\,$ ionizing photons s$^{-1}$ was turned on at the densest cell, in 
a virialized halo
of total mass $10^{11}\,M_\odot$. The box length is 2.4 comoving
Mpc. The solid contours give the position of the front at 0.15, 0.25, 0.38, and
0.57\,Myr after the quasar has switched on at $z=7$. The underlying greyscale
image indicates the initial \HI density field. (From Abel \etal 1999.)
}
\end{figure}

Since the mean free path of Lyc radiation is always much 
smaller than the horizon (this is also true after `overlapping' because of 
the residual \HI still present in the \lya forest clouds and the Lyman-limit 
systems), the filling factor of cosmological \HII regions is 
equal at any given time $t$ to the total number of ionizing photons 
emitted per hydrogen atom by all radiation sources present at earlier epochs,
$\int_0^t \dot n_{\rm ion}dt'/\bar{n}_\nH$, minus the total number of radiative
recombinations per atom, $\int_0^t Q dt'/\bar{t}_{\rm rec}$. This
statement reflects the simple fact that {\it every ultraviolet photon that is
emitted is either absorbed by a newly ionized hydrogen atom or by a 
recombining one.}
Differentiating one gets
$$
\frac{dQ}{dt}=\frac{\dot n_{\rm ion}}{\bar{n}_\nH}-\frac{Q}
{\bar{t}_{\rm rec}}  \eqno(5)
$$
(Madau, Haardt, \& Rees 1999).
It is this differential equation -- and its equivalent for
expanding helium zones -- that statistically describes the transition
from a neutral Universe to a fully ionized one independently, for a 
given UV photon emissivity per unit cosmological volume $\dot n_{\rm ion}$, 
of the complex and possibly short-lived emission histories of
individual radiation sources, e.g. on whether their comoving space density is
constant or varies with cosmic time. Initially, when the filling factor is 
$\ll 1$, recombinations can be neglected and the ionized volume increases 
at a rate fixed solely by the ratio $\dot n_{\rm ion}/\bar{n}_\nH$. 
As time goes on and more and more Lyc photons are emitted, radiative 
recombinations become important and slow down the growth of the ionized 
volume, until $Q$ reaches unity, the recombination term saturates, and 
reionization is finally completed (except for the high density regions 
far from any source which are only gradually eaten away, Miralda-Escud\'{e}, 
Haehnelt, \& Rees 2000). In the limit of a fast recombining IGM 
($\bar{t}_{\rm rec}\ll t$), one can neglect the derivative on the left-hand 
side of equation (5) and derive   
$$
Q\lta \frac{\dot n_{\rm ion}}{\bar{n}_\nH}\bar{t}_{\rm rec},
\eqno(6)
$$
i.e. the volume filling factor of ionized bubbles must be less (or equal) to the
number of Lyc photons emitted per hydrogen atom in one recombination time. 
In other words, because of radiative recombinations, only a fraction 
$\bar{t}_{\rm rec}/t\ll 1$ of the photons emitted above 1 ryd is actually used 
to ionize new IGM material. The Universe is completely reionized when 
$$
\dot n_{\rm ion} \bar{t}_{\rm rec}\gta \bar{n}_\nH,
\eqno(7)
$$
i.e. when emission rate of ultraviolet photons exceeds the mean rate of
recombinations.
 
\subsection{A clumpy Universe}

\ni The simplest way to treat reionization in a inhomogeneous medium is in 
terms of a clumping factor that increases the effective gas recombination rate. 
In this case the volume-averaged recombination time is
$$
\bar{t}_{\rm rec}=[(1+2\chi) \bar{n}_p\alpha_B\,C]^{-1}=0.06\, {\rm Gyr}
\left(\frac{\Omega_B h^2}{0.02}\right)^{-1}\left(\frac{1+z}{10}\right)^{-3}
\frac{\bar n_\nH}{\bar n_p}~C_{10}^{-1}, \eqno(8)
$$
where $\alpha_B$ is the recombination coefficient to the excited states of 
hydrogen (at an assumed gas temperature of $10^4\,$K), $\chi$ the helium to 
hydrogen abundance ratio, and the factor $C\equiv \langle n_p^2\rangle/
\bar{n}_p^2>1$ takes into account the degree of clumpiness of photoionized 
regions (hereafter $C_{10}\equiv C/10$). If ionized gas with density $n_p$ 
filled uniformly a fraction $1/C$ 
of the available volume, the rest being empty space, the mean square density 
would be $\langle n_p^2\rangle =n_p^2/C=\bar{n}_p^2C$. More in general, if 
$f_m$ is the fraction of baryonic mass in photoionized 
gas at an overdensity $\delta$ relative to the mean, and the remaining 
(underdense) medium is distributed uniformly, then the fractional volume 
occupied by the denser component is   
$$
f_v=f_m/\delta, \eqno(9) 
$$
the density of the diffuse component is
$$
\bar{n}_p\frac{1-f_m}{1-f_v}, \eqno(10) 
$$
and the recombination rate is larger than that of a homogeneous Universe by
the factor 
$$
C=f_m\delta+\frac{(1-f_m)^2}{1-f_v} \eqno(11)
$$  
(e.g. Chiu \& Ostriker 1999; Valageas \& Silk 1999). 
It is difficult to estimate the clumping factor accurately. According to 
hydrodynamics simulations of structure formation in the IGM (within the 
framework of 
CDM-dominated cosmologies), \Lya forest clouds with moderate overdensities, 
$5\lta\delta\lta 10$, occupy a fraction of the available volume which
is too small for them to dominate the clumping at high
redshifts (e.g. Zhang \etal 1998; Theuns \etal 1998). In hierarchical 
clustering models, it is the virialized gas (with $\delta\approx 180$ if one 
ignores the slope of the density profile) in dark matter halos with 
temperatures $\lta 10^{4}\,$K (masses $M\lta 10^7\,h^{-1}\,M_\odot$) which 
will plausibly boost the recombination rate by large factors as soon as the 
collapsed mass fraction exceeds 0.5\%. Halos or
halo cores which are dense and thick enough to be self-shielded from UV 
radiation will stay neutral 
and will not contribute to the recombination rate. This is also true of gas 
in more massive halos, which will be virialized to higher temperatures 
and ionized by collisions with thermal electrons. 
With a large comoving space density at 
$z=9$ of $dn/d\ln M\sim 1000\,h^3\,$Mpc$^{-3}$, corresponding to a mean 
proper distance of only $\sim 6\,h^{-1}\,$kpc, and to a mass fraction of 
$0.04$, halos with $T_v\approx 10^4\,$K will contribute significantly, 
$f_m\delta\approx 7$, to the clumping.     
Recent calculations by Benson \etal (2000), which instead include 
all halos with $T_v>10^4\,$K and adopt an isothermal density profile with a 
flat core, give $C\approx 30$ already at $z=9$. 
Because of finite resolution effects, numerical simulations
may underestimate clumping: in those of Gnedin \& Ostriker (1997), 
for example, $C$ rises above unity at $z\lta 20$, and grows to $C\sim 10$ 
(40) at $z\approx 9$ (5). 

It is important to note that the use of the volume-averaged clumping factor 
in the recombination timescale is
only justified when the size of the \HII regions is much larger compared to the
scale of the clumping, so that the effect of many halos within
the ionized volume can be averaged over. This will be a good approximation
either at late epochs, when the \HII zones have 
had time to grow (or when overlapping ionized regions from an ensemble of 
sources are able to proper sample the small-scale density fluctuations), or at
earlier epochs if
the ionized bubbles are produced by more luminous sources like quasars
or the stars within halos collapsing from high-$\sigma$ peaks.
As mentioned above, the mean free path between halos having $T_v\approx 10^4\,$K
is $\lambda\sim 6\,h^{-1}\,$kpc at $z=9$, but their virial radius is only 
$r_v\approx 0.4\,h^{-1}\,$kpc. It is only on scales greater than 
$\lambda^3/r_v^2\approx 2\,h^{-1}\,$Mpc that the clumping can then be 
averaged over, and the covering factor of halos within the Str\"omgren sphere 
exceeds unity. 

\section{Sources of UV photons}

\subsection{Quasars}

\ni In recent years, several optical surveys (Warren, Hewett, \& Osmer 1994; 
Schmidt, Schneider, \& Gunn 1995; Kennefick, Djorgovski, \& de Carvalho 1995) 
have consistently provided 
evidence for a turnover in the QSO counts. The space density of radio-loud 
quasars also appears to decrease strongly for $z>3$ (Shaver \etal 1996), 
suggesting that the turnover is indeed real and not an effect on 
optically-selected QSOs induced by dust along the line of sight.
The density of optically bright and flat-spectrum radio-loud quasars 
has a relatively flat maximum at $1.8\lta z\lta 2.8$, and declines gradually
at higher redshifts (Figure 2). 

\begin{figure}
\epsfysize=12cm 
\epsfxsize=12cm 
\hspace{1.cm}\epsfbox{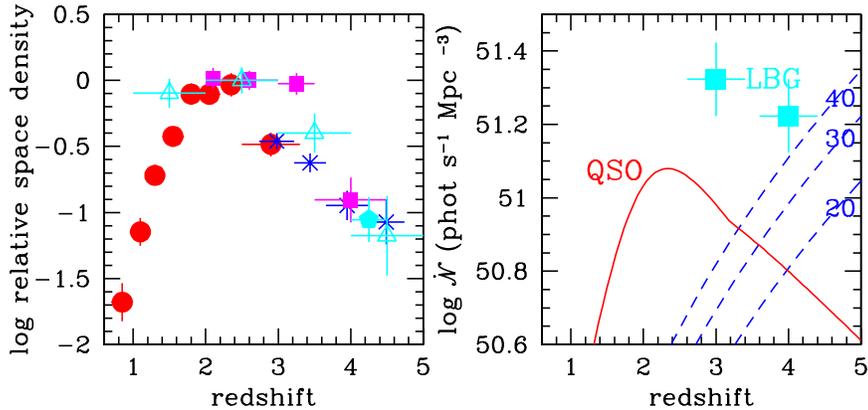}
\vspace{-5.5cm}
\caption{
{\it Left}: comoving space density of bright QSOs as a
function of redshift. The data points with error bars are taken from
different optical and radio surveys (see Madau, Haardt, \& Rees 1999
for details). {\it Right}: comoving emission rate of hydrogen Lyc 
photons ({\it solid line}) from QSOs, compared to the comoving rate of
recombinations, $\bar{n}_\nH(0)/\bar{t}_{\rm rec}$ ({\it dashed lines}) in 
a IGM with gas clumping factor $C=20, 30, 40$). 
A EdS cosmology with $\Omega_Bh^2=0.02$ and $h=0.5$ has been assumed. 
Models based on photoionization by quasar sources appear to fall short at 
$z\approx 5$. The data points show the estimated contribution from Lyman-break
galaxies at $z\approx 3$ and 4, assuming that the fraction of Lyc photons 
which escapes the dense \HI layers into the galaxy halos and the IGM is 50\%.
}
\end{figure}

The QSO emission rate of hydrogen Lyc photons per unit comoving volume,
$\dot{\cal N}_Q$, is also shown in Figure 2. The procedure adopted to 
derive this quantity implies a large
correction for incompleteness at high-$z$. With a fit to the quasar
luminosity function (LF) which goes as $\phi(L)\propto L^{-\beta}$, with 
$\beta=1.64$ at the faint end (Pei 1995), the
contribution to the emissivity converges rather slowly, as $L^{0.36}$. At
$z=4$, for example, the blue magnitude at the break of the LF is $M_*\approx
-25.4$, comparable or slightly fainter than the limit of current high-$z$
QSO surveys. While a large fraction, about 90\% at $z=4$ and even higher at
earlier epochs, of the ionizing emissivity shown in the figure is therefore
produced by quasars that have not been actually observed, and are
assumed to be present based on an extrapolation from lower redshifts, it
is also fair to ask whether an excess of low-luminosity
QSOs, relative to the best-fit LF, could actually boost the estimated Lyc
emissivity at early epochs. The interest in models where the quasar LF
significantly steepens with lookback time, and therefore predict many more
QSOs at faint magnitudes than the extrapolation of Pei's (1995) fitting
functions, stems from recent claims
of a strong linear correlation between bulge and observed black hole masses
(Magorrian \etal 1998), linked to the steep mass function of dark matter haloes
predicted by hierarchical cosmogonies (e.g. Haehnelt, Natarajan, \& Rees 1998;
Haiman \& Loeb 1998).
As discussed by Haiman, Madau, \& Loeb (1999), the space density of 
low-luminosity quasars at high-$z$ is constrained by the observed lack of 
red, unresolved faint objects in the {\it Hubble Deep Field} (HDF). 
Down to a 50\%
completeness limit of $V_\AB=29.6$ ($I_\AB=28.6$), no $z>4$ quasar candidates
have actually been found by Conti \etal (1999): by contrast, about 10 objects 
would be predicted by a QSO evolution model characterized by a steep LF with 
slope $\beta=2$  and a comoving space 
density that remains constant above $z=2.5$ instead of dropping (Figure 3),
and choosen to boost the emission rate of ultraviolet photons at $z\sim 5$ 
by a factor of 5. A large population of faint AGNs at high-$z$ would still be
consistent with the data if, at these faint magnitude levels and high image
resolution, the host galaxies of active nuclei could actually be resolved
by the {\it Hubble Space Telescope}.

\begin{figure}
\epsfysize=8cm 
\epsfxsize=8cm 
\hspace{2.5cm}\epsfbox{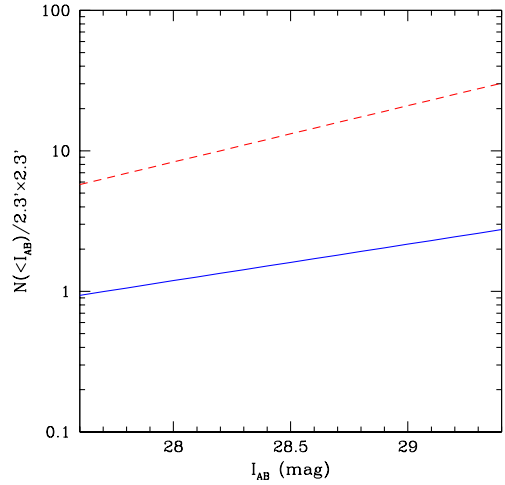}
\vspace{0.cm}
\caption{Theoretical number-magnitude relation of quasars in the redshift range
$4.0<z<5.5$. The {\it solid line} shows the prediction for a `standard' 
QSO model,
one in which the faint end of the QSO luminosity function has slope
$\beta=1.64$ and tracks the turnover observed in the space density of bright
quasars at $z\gta 3$. The {\it dashed line} shows the increased number of
sources expected in the case of a steeper, $\beta=2$, luminosity function
and a comoving space density that stays constant above $z=2.5$. The latter
evolution scenario provides, within the errors, enough UV photons to keep the
Universe ionized at $z\approx 5$, but appears to be inconsistent with the lack
of red, faint stellar objects observed in the {\it Hubble Deep Field}.
}
\end{figure}

\subsection{Star-forming galaxies}

\ni Galaxies with ongoing star-formation are another obvious source of Lyc 
photons. The recent progress in our understanding of faint galaxy
data made possible by the identification of star-forming galaxies at
$2\lta z\lta 4$ in ground-based surveys and in the HDF has provided new
clues to the long-standing issue of whether galaxies at high redshifts can
provide a significant contribution to the ionizing background flux. Since the
rest-frame UV continuum at 1500 \AA\ (redshifted into the visible band for a
source at $z\approx 3$) is dominated by the same short-lived, massive stars
which are responsible for the emission of photons shortward of the Lyman edge,
the needed conversion factor, about one Lyc photon every 10 photons at
1500 \AA, is fairly insensitive to the assumed IMF and is independent of the
galaxy history for $t\gg 10^{7.3}\,$ yr.

Composite ultraviolet luminosity functions of Lyman-break galaxies (LBG) at
$z\approx 3$ and $z\approx 4$ have been recently derived by Steidel \etal 
(1999). They are based on a large catalog of spectroscopically and 
photometrically selected galaxies from the ground-based and HDF samples, and 
span about a factor of 40 in luminosity from the faint to the bright end. 
Integrating these LF over all luminosities $L>0.1L^*$, 
and using the conversion $L(1500)/L(912)\approx 6$ valid for a Salpeter
mass function and constant star formation rate, we derive
for the comoving emissivities at 1 ryd the values of
$9\pm 2\times 10^{25}\,h\,\emunits$ at $z\approx 3$, and  
$7\pm 2\times 10^{25}\,h\,\emunits$ at $z\approx 4$, about 4 times higher 
than the estimated quasar contribution at $z=3$. These numbers do not include
any correction for local \HI absorption (since the color excess 
$E_{912-1500}$ is expected to be small, dust exinction can probably be neglected
in correcting from observed rest-frame far-UV to the Lyman edge). 
The data points plotted in Figure 2 assumes a value of $f_{\rm esc}=0.5$ for 
the unknown fraction of Lyc photons which escapes the dense sites
of star formation (not included in our clumping factor) into the halos and
the intergalalactic space. 
Note that, at $z=3$, Lyman-break galaxies radiate more ionizing photons than 
QSOs for $f_{\rm esc}\gta 25\%$.

\section{Implications}

\subsection{First light}

\ni We have seen in the previous sections that, in the approximation the 
clumping
can be averaged over, only the photons emitted within one recombination
timescale can actually be used to ionize new material. As $\bar{t}_{\rm rec}
\ll t$ at high redshifts, it is possible to compute using equation (7) a
critical value for the photon emission rate per unit cosmological comoving
volume at a given epoch, $\dot {\cal N}_c$,
independently of the (unknown) previous emission 
history of the Universe: only rates above this value will provide enough UV 
photons to keep the IGM ionized at that epoch. 
Equation (7) can then be rewritten as
$$
\dot {\cal N}_c(z)=\frac{\bar{n}_\nH(0)}{\bar{t}_{\rm rec}(z)}=
(10^{51.4} \, \ndotunits)\, C_{10} \left(\frac{1+z}{10}\right)^{3}
\left(\frac{\Omega_B h^2}{0.02}\right)^2. \eqno(12)
$$
The uncertainty on this value is difficult to estimate, as it depends
on the clumping factor and the nucleosynthesis constrained baryon 
density. It is interesting to convert this rate into a `minimum' star 
formation rate per unit (comoving) volume, $\dot \rho_*$ (for $\Omega_Bh^2=
0.02$):
$$
{\dot \rho_*}=\frac{\dot {\cal N}_c\times 10^{-53.1}}{f_{\rm esc}}
\approx (0.12\,\sfrd)\,\left(\frac{0.5}{f_{\rm esc}}\right)\,C_{10} 
\left(\frac{1+z}{10}\right)^{3}. \eqno(13) 
$$
(The conversion factor can be understood by noting that, for each 1 $M_\odot$ 
of stars formed,
8\% goes into massive stars with $M>20 M_\odot$ that dominate the
Lyc luminosity of a stellar population. At the end of the C-burning
phase, roughly half of the initial mass is converted into helium and carbon,
with a mass fraction released as radiation of 0.007. About 25\% of the energy
radiated away goes
into Lyc photons of mean energy 20 eV. For each 1 $M_\odot$ of stars
formed every year, we then expect $0.08\times 0.5 \times 0.007 \times 
0.25\times M_\odot c^2/20 {\,\rm eV\, yr} \sim 10^{53}\ndotun$
to be emitted shortward of 1 ryd.) 

Taken at face value, equations (12) and (13) have perhaps a surprising 
implication. In a inhomogeneous Universe, early reionization at $z\sim 9$ 
requires an ionizing emissivity which is {\it comparable or larger} than that 
radiated by QSOs at the peak of their activity, $z\approx 3$. In a similar
manner,  
photoionization by massive stars can only play a role if the star formation 
density at this epoch were significantly larger than the value directly 
`observed' (i.e. uncorrected for dust reddening) at $z=2$ (Madau, 
Pozzetti, \& Dickinson 1998).  

\begin{figure}
\epsfysize=9cm 
\epsfxsize=9cm 
\hspace{1.5cm}\epsfbox{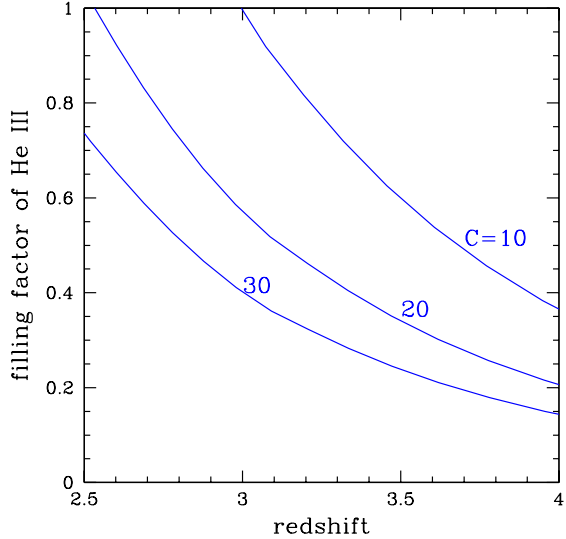}
\vspace{0.0cm}
\caption{The evolution of the \HeIII filling factor as a  function of redshift 
in a inhomogeneous Universe where photoionization is dominated by QSOs
turning over at $z\gta 3$. 
From right to left, the three curves assume a constant clumping factor of 
$C=10, 20,$ and 30. The QSO photon spectrum is assumed to vary as $\nu^{-2.8}$ 
shortward of the hydrogen Lyman edge. Note how the ionization of \HeII is 
never completed before $z=3$ in models with $C\ge 10$.
}
\end{figure}

\subsection{Delayed \HeII reionization}

\ni Because of its higher ionization potential and the steep spectra of UV
radiation sources, the most abundant (by a factor $\sim 100$) absorbing ion in 
the post-reionization Universe is not \HI but \HeII. The
importance of intergalactic helium in the context of this study stems from the
possibility of detecting the effect of `incomplete' \HeII reionization in the
spectra of $z\sim 3$ quasars as, depending on the clumpiness of the IGM (Madau
\& Meiksin 1994), the photoionization of  singly ionized helium may be delayed
until much later than for \HI.

Since \HI and \HeI do not absorb a significant fraction of $h\nu>54.4\,$ eV
photons, the problem of \HeII reionization can be decoupled from that of
other ionizations, and the equivalent of equation (5) for expanding \HeIII
regions becomes
$$
\frac{dQ}{dt}=\frac{\dot n_{\rm ion4}}{\bar{n}_\nHe}-\frac{Q}{\bar{t}_\nHeIII}, \eqno(14)
$$
where $\dot n_{\rm ion4}$ now includes only photons above 4 ryd,
and $\bar{t}_\nHeIII$ is $6.5$ times shorter than the hydrogen
recombination timescale if ionized hydrogen and doubly ionized helium have
similar clumping factors.\footnote{This last assumption appears, however,
rather dubious:
the reason is that self-shielding of \HeII Lyc radiation occurs at 
much lower hydrogen columns than self-shielding of photons at $1\,$ryd 
(by about a factor of $S/2$, where the spectral `softness' $S$ is the 
the ratio of the radiation flux at the hydrogen Lyman edge to the flux 
at 4 ryd), and self-shielded gas will remain neutral and not add to 
the recombination
rate. Ionized hydrogen may then be more clumpy than doubly ionized helium.}~
It is interesting to note that, if the intrinsic photon spectrum of ionizing 
sources has slope $\dot n(\nu)\propto \nu^{-2.8}$, the first
terms on the right-hand side of equations (5) and (14)
are actually equal, and a significant delay between the complete overlapping
of \HII and \HeIII regions can only arise if recombinations are important.
This effect is illustrated in Figure 4, where the expected evolution of the
\HeIII filling factor (obtained by numerical integration of eq. 14) is plotted
for a QSO-photoionization model with a source decline at high redshifts:
\HeII reionization is never completed before $z=3$ in models with $C\gta 10$.
A significant contribution to the UV background at 4 ryd from massive stars,
which could push the helium reionization epoch to higher redshifts, has been 
traditionally ruled out on the basis that the ratio between the number of 
\HeII and \HI Lyc photons emitted from low-metallicity starbursts is only 
about two percent (Leitherer \& Heckman 1995), five times smaller than 
in typical QSO spectra. It has been recently pointed out by Tumlinson \& Shull
(2000), however, that metal-free stars exhibit higher effective temperatures 
and dramatically harder stellar spectra, particularly in the \HeII continuum.
This enhanced He-ionizing capabilities of Pop III stars could have interesting
implications for reionization.    

To date, various studies of the HeII \lya forest in the spectra of distant QSOs 
(Hogan, Anderson, \& Rugers 1997; Reimers \etal 1997; Heap \etal 2000)
have revealed patchy absorption with low \HeII opacity `voids' alternating 
several Mpc sized regions with vanishing flux. These observations suggest
that helium
absorption does not increase smoothly with lookback time, but rather in
the abrupt manner expected in the final stages of inhomogeneous 
reionization by quasar sources. Radiative transfer effects during 
\HeII reionization could affect the thermal history of the IGM (Abel \& 
Haehnelt 1999, Efstathiou, this volume). Here it is important to remark that, while delayed
\HeII reionization in a clumpy Universe appears to be naturally linked to the
observed decline in the space density of quasars beyond $z\sim 3$, the
complete overlapping of \HeIII regions occurs instead much earlier ($z\gg 5$)
in models that predict many more faint QSOs at high redshifts (Haiman \& Loeb
1998).


\bigskip\bigskip
\ni I would like to thank my collaborators, T. Abel, F. Haardt, Z. Haiman, and 
M. Rees, for many useful discussions on the topics discussed here.

\end{document}